\documentclass[letterpaper,10pt]{IEEEtran}

\usepackage{amsmath}
\usepackage{amssymb}
\usepackage{setspace}
\usepackage{epsfig}
\usepackage{balance}
\usepackage[noadjust]{cite}
\usepackage{flushend}
\usepackage{algorithm}
\usepackage{algorithmic}

\title{Near Maximum-Likelihood Performance of Some New Cyclic Codes Constructed in the Finite-Field Transform Domain}

\IEEEoverridecommandlockouts

\author{
\authorblockN{C. Tjhai, M. Tomlinson, R. Horan, M. Ambroze and M. Ahmed\\}
\authorblockA{Fixed and Mobile Communications Research,\\University of Plymouth,\\Plymouth PL4 8AA,\\United Kingdom,\\
              email: \{ctjhai,mtomlinson,rhoran,mambroze,mahmed\}@plymouth.ac.uk}
\thanks{This work was partly funded by the Overseas Research Students award.}
}

\begin{document}

\maketitle

\begin{abstract}
It is shown that some well-known and some new cyclic codes with orthogonal parity-check equations
can be constructed in the finite-field transform domain. It is also shown that, for
some binary linear cyclic codes, the performance of the iterative decoder can be improved by
substituting some of the dual code codewords in the parity-check matrix with other dual code codewords
formed from linear combinations.
This technique can bring the performance of a code closer to its maximum-likelihood
performance, which can be derived from the erroneous decoded codeword whose euclidean distance with
the respect to the received block is smaller than that of the correct codeword.
For $(63,37)$, $(93,47)$ and $(105,53)$ cyclic codes, the maximum-likelihood performance is realised
with this technique.
\end{abstract}


\section{Introduction}\label{sec:Intro}

Low-density parity-check (LDPC) codes~\cite{Gallager.1962},\cite{MacKay_et_al.1996} form a class of
$(n, k)$ linear block codes, where $n$ is the codeword length and $k$ is the information length,
that can approach near capacity performance. The good performance of LDPC
codes is attributed to the code representation and the use of an iterative decoder.
It is an essential condition
that the code representation does not contain more than one parity-check equation checking on the same
two or more bit positions, i.e. no cycles of length 4. The avoidance of these short cycles is
important to allow convergence of the iterative decoder~\cite{Chen_et_al.2002}. The best performance gains
to date have been obtained with long LDPC codes, i.e. several thousand bits in length.

There are many applications where short LDPC codes can be potentially useful. Applications such as
watermarking, thin data storage, command/control data reporting and packet communications require blocks
of data ranging from 32 to 512
bits to be either robustly protected or reliably transmitted. We concentrate on cyclic LDPC codes of
similar lengths in this paper. The particular class of cyclic codes we consider are difference set
cyclic (DSC) codes~\cite{Weldon.1966} and one-step majority logic decodable (OSMLD)
codes~\cite{Lin_et_al.1983} which have orthogonal parity-check equations on
each bit position, thus there are no cycles of length 4. For short block lengths, these cyclic
codes\footnote{We will refer the OSMLD and DSC codes as cyclic codes from this point onwards}
have been shown to outperform the ad-hoc computer design (random) counterpart of the same code-rate
and block length~\cite{Lucas_et_al.2000}. For an $(n, k)$ LDPC code, the ad-hoc
computer design code has $n-k$ parity-check equations but the cyclic code has $n$ parity-check
equations that can be used by the iterative decoder. Consequently, cyclic codes exhibit better
convergence than the random LDPC codes when iteratively decoded.

In this paper, we present a modified Belief-Propagation (BP) iterative decoder that can perform
near maximum-likelihood (ML) performance for binary transmission over the additive-white-Gaussian-noise (AWGN) channel.
It is also shown that, for certain cyclic codes, the modified iterative decoder can achieve ML performance.

The organisation of this paper is as follows. Section~\ref{sec:fft-cyclic} gives a brief
review on cyclic code construction method in the finite-field transform domain.
Section~\ref{sec:mrl} introduces the idea of a more-likely codeword and its relationship to ML and
iterative decoders. We present modification to the iterative decoder in section~\ref{sec:mod_iterative_dec}
and some simulation results of the modified decoder are presented in section~\ref{sec:results}.
Section \ref{sec:conclusion} contains the conclusions.


\begin{table*}[!t]
\renewcommand{\arraystretch}{1.3}
\caption{\label{tbl:osmld_codes}Examples of the constructed cyclic codes}
\centering
\begin{tabular}{c|p{4.5in}|c}\hline
$(n,k)$ & \multicolumn{1}{|c|}{$u(x)$} & $d_{min}$\\\hline
$(21,11)$  & $1+x^2+x^7+x^8+x^{11}$ & 6 \\\hline
$(63,37)$  & $1+x^1+x^3+x^7+x^{15}+x^{20}+x^{31}+x^{41}$ & 9\\\hline
$(73,45)$  & $1+x+x^3+x^7+x^{15}+x^{31}+x^{36}+x^{54}+x^{63}$ & 10\\\hline
$(93,47)$  & $1+x^3+x^9+x^{21}+x^{28}+x^{45}+x^{59}$ & 8\\\hline
$(105,53)$ & $1+x^7+x^8+x^{21}+x^{23}+x^{49}+x^{53}$ & 8\\\hline
$(255,175)$ & $1+x+x^3+x^7+x^{15}+x^{26}+x^{31}+x^{53}+x^{63}+x^{98}+x^{107}+x^{127}+x^{140}+x^{176}+x^{197}+
x^{215}$ & 17\\\hline
$(341,205)$& $1+x^1+x^3+x^7+x^{15}+x^{31}+x^{54}+x^{63}+x^{98}+x^{109}+x^{127}+x^{170}+x^{197}+x^{219}+x^{255}$ & 16\\\hline
$(511,199)$ & $1+x+x^3+x^7+x^{15}+x^{31}+x^{63}+x^{82}+x^{100}+x^{127}+x^{152}+x^{165}+x^{201}+x^{255}+x^{296}+
x^{305}+x^{331}+x^{403}$ & 19\\\hline
$(511,259)$ & $1+x^{31}+x^{42}+x^{93}+x^{115}+x^{217}+x^{240}+x^{261}+x^{360}+x^{420}+x^{450}+x^{465}$ & 13\\\hline
\end{tabular}
\end{table*}

\section{Finite-Field Transform Domain Construction of Binary Cyclic Codes}\label{sec:fft-cyclic}

There are relatively few OSMLD and DSC codes. As shown in~\cite{Horan_et_al.proc-iee-comms}, we have
extended these codes by a construction method that works in the finite-field transform domain, which
is also known as the Mattson-Solomon domain. We briefly review the construction method in this section.

Let $n$ be a positive odd integer and $\text{GF}(2^m)$ be the splitting field for $1+x^n$ over
$\text{GF}(2)$. We assume that $\alpha$ is the generator for $\text{GF}(2^m)$ and $T_a(x)$ is the polynomial
with coefficients in $\text{GF}(2^a)$ and degree $\le n-1$. Let us denote
$\mathcal{F}=\{f_1(z),f_2(z),\ldots,f_t(z)\}$, where $f_i(z) \in T_1(z)$ is an irreducible polynomial,
such that $\prod_{i\le i \le t} f_i(z) = 1+z^n$. For each $f_i(z)$, there is a corresponding primitive
idempotent\footnote{A binary polynomial, $e(x)$, is an idempotent if the property of
$e(x) = e(x)^2 = e(x^2)\text{ mod }1+x^n$ is satisfied.}, denoted as $\theta_i(z)$, which can be obtained as
follows:
\begin{align}
\theta_i(z) &= \frac{z(1+z^n)f^\prime_i(z)}{f_i(z)} + \delta(1+z^n)\label{eqn:theta_z}
\end{align}
where $f^\prime_i(z) = \frac{d}{dz} f_i(z)$, $f^\prime_i(z) \in T_1(z)$ and the integer $\delta$ is:
\begin{align*}
\delta &=
\begin{cases}
1 & \text{if deg}(f_i(z))\text{ is odd,}\\
0 & \text{otherwise.}
\end{cases}
\end{align*}
where $\text{deg}(a(x))$ represents the degree of the polynomial $a(x)$.

Let $a(x) \in T_m(x)$, the finite-field transform or Mattson-Solomon (MS) polynomial of $a(x)$ is:
\begin{align}
A(z) &= \text{MS}(a(x)) = \sum_{j=0}^{n-1}a(\alpha^{-j})z^j\label{eqn:forward_ms}\\
a(x) &= \text{MS}^{-1}(A(z)) = \frac{1}{n}\sum_{i=0}^{n-1}A(\alpha^i)x^i\label{eqn:inverse_ms}
\end{align}
where $A(z) \in T_m(z)$. 

Let $\mathcal{I} \subseteq \{1,2,\ldots,t\}$, we define $f(z) = \prod_{i \in \mathcal{I}} f_i(z)$
and $\theta(z) = \sum_{i \in \mathcal{I}} \theta_i(z)$, where $f(z),\theta(z) \in T_1(z)$. Let
us define a binary polynomial $u(x) = \text{MS}(\theta(z))$. Since the MS polynomial of a binary
polynomial is an idempotent and vice-versa~\cite{MacWilliams_et_al.1977}, $u(x)$ is an idempotent
with coefficients in $\text{GF}(2)$. If we write $u(x) = u_0 + u_1x + \ldots + u_{n-1}x^{n-1}$
then, from equation~\ref{eqn:inverse_ms}
\begin{align}
u_i = \frac{1}{n} \theta(\alpha^i),\quad\forall i \in \{0,1,\ldots,n-1\}.\label{eqn:ui}
\end{align}
The idempotent $u(x)$ can be used to describe an $(n,k)$ binary cyclic code which has a parity-check
polynomial, $h(x)$, of degree $k$ and a generator polynomial, $g(x)$, of degree $n-k$. The polynomial
$h(x)$ is a divisor of the idempotent $u(x)$, i.e. $(u(x), 1+x^n)=h(x)$\footnote{where $(a,b)$ denotes
the greatest common divisor of $a$ and $b$.} and $u(x) = m(x)h(x)$ where $m(x)$ contains the repeated
factors and/or non-factors of $1+x^n$.

From the theories above, we can summarise that:
\begin{enumerate}
\item[1)] The weight of $u(x)$ is equal to the number of $n$th roots of unity which are roots of $f(z)$.
Note that for $0 \le i \le n-1$, $\theta(\alpha^i)=1$ if and only if $f(\alpha^i) = 0$ and
from equation~\ref{eqn:ui}, $u_i = 1$ if and only if $\theta(\alpha^i) = 1$.
In the other words, $u_i = 1$ precisely when $f(\alpha^i) = 0$, giving $\text{wt}(u(x))$\footnote{
$\text{wt}(a(x))$ denotes the weight of polynomial $a(x)$.} $= \text{deg}(f(z))$.
Clearly, $\text{wt}(u(x)) = 
\sum_{i \in \mathcal{I}} \text{deg}(f_i(z))$.
\item[2)] Since $\theta(z) = \text{MS}(u(x))$, the number of zeros of $u(x)$ which are roots of unity
is clearly $n - \text{wt}(\theta(z))$.
\end{enumerate}

In general, $\text{wt}(u(x))$ is much lower than $\text{wt}(h(x))$ and as such, we can
derive a low-density parity-check matrix from $u(x)$ and apply iterative decoding on it. The parity-check
matrix of the resulting code consists of the $n$ cyclic shifts of $x^nu(x^{-1})$. Since the idempotent
$u(x)$ is orthogonal on each bit position, the resulting LDPC code has no cycles of length 4 in the
bipartite graph and the true minimum-distance, $d_{min}$, of the code is simply $\text{wt}(1+u(x))$,
see~\cite[Theorem 10.1]{Peterson_et_al.1972} for the proof.

Table~\ref{tbl:osmld_codes} shows some examples of cyclic codes derived from this technique. From
Table~\ref{tbl:osmld_codes}, it is clear that our technique can also be used to construct the well-known
OSMLD and DSC codes.


\section{More Likely Codewords in Relation to ML and Iterative Decoders}\label{sec:mrl}

Realising an optimum decoder for any coded system is NP-complete~\cite{Berlekamp_et_al.1978}.
For general $(n, k)$ binary linear codes, the optimum decoding complexity is proportional to
$\text{min}\{2^{k}, 2^{n-k}\}$. Due to this complexity, the optimum decoder can only be realised
for very short or very high-rate or very low-rate codes.
An ML decoder is the optimum decoder in terms of minimising the frame-error-rate (FER).
An ML decoder will output a codeword that has the closest euclidean distance~\cite{Lin_et_al.1983}
to the received block. 

\begin{figure}[hbt]
\centering
\includegraphics[width=2in]{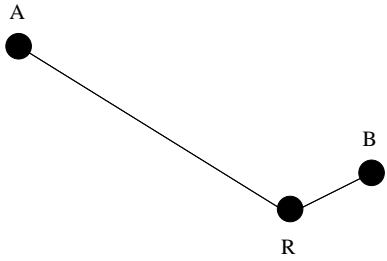}
\caption{\label{fig:mrl}ML decision criterion}
\end{figure}

The iterative decoder is a suboptimal decoder approximating ML performance. In
decoding LDPC codes, the BP iterative decoder can produce a codeword that is not identical
to the transmitted codeword. This is illustrated by the two-dimensional representation of the
ML decision criterion shown in Fig.~\ref{fig:mrl}.
Points \textsf{R} and \textsf{A} represent the received block
and transmitted/correct codeword respectively. The point \textsf{B} represents a codeword whose
euclidean distance with the respect to \textsf{R} is smaller than that of \textsf{A}.
If the iterative decoder outputs codeword \textsf{B} then a decoding error
is produced, but an ML decoder will also make an error. We classify
codeword \textsf{B} as a more likely (mrl) codeword~\cite{Ambroze_et_al.2003},\cite{Papagiannis_et_al.2004}.
By counting the number of mrl codewords produced in a simulation, we can derive an mrl-FER curve.
A similar technique has been used by Dorsch~\cite{Dorsch.1974}, but the metric was based on the hamming
distance rather than the euclidean distance, i.e. hard-decisions rather than soft-decisions.

The significance of the mrl codewords is that an ML decoder either outputs correct codewords or mrl
codewords. The percentage of mrl codewords output from the iterative decoder gives us a performance
indication of how close the iterative decoder is from the ML decoder for the same code. The mrl-FER
provides the lower-bound on the ML performance of a code in comparison to the Maximum-Likelihood-Asymptote
(MLA) which provides the upper-bound.


\section{Improved Belief-Propagation Decoder}\label{sec:mod_iterative_dec}


For the $(63,37)$ cyclic code, it has been noticed that the standard BP decoder produces many codewords
that are neither correct nor mrl.  The number of incorrect codewords is much larger than the number of
mrl codewords output. Based on these findings and the
fact that every codeword will satisfy all $2^{n-k}$ parity-check equations,
we should be able to improve the performance of the BP decoder by
extending the number of parity-check equations in the parity-check matrix, denoted as $\mathrm{\textbf{H}}$.
However, this is likely to be true if the extended parity-check matrix have low-density and does not
contain many short cycles.

For any linear codes, additional parity-check equations can be formed from the linear combinations of the
equations in $\mathrm{\textbf{H}}$. These additional parity-check equations form the high weight codewords
of the dual code and appending them to $\mathrm{\textbf{H}}$ will introduce many short cycles.

The proposed modified BP decoder does not extend the number of parity-check equations in
$\mathrm{\textbf{H}}$. Instead, we generate a set of parity-check equations, denoted as
$\mathrm{\textbf{H}^e}$, by taking the linear combinations of those equations in $\mathrm{\textbf{H}}$.
A subset of $\mathrm{\textbf{H}^e}$ is substituted into $\mathrm{\textbf{H}}$ resulting in a modified parity-check
matrix, labelled as $\hat{\mathrm{\textbf{H}}}$.
The overall procedures is described in Algorithm~\ref{algo:dec}~\cite{Tjhai_et_al.patent}.
Note that the selection of the parity-check equations may be made on a random basis or may correspond to a
predetermined sequence.

\renewcommand{\algorithmicrequire}{\textbf{Input:}}
\renewcommand{\algorithmicensure}{\textbf{Output:}}
\begin{algorithm}
\caption{Modified Belief-Propagation Iterative Decoder}\label{algo:dec}
\begin{algorithmic}[1]
\REQUIRE$\text{ }$\\
$\mathbf{r} \Leftarrow$ received vector\\
$\mathrm{\textbf{H}} \Leftarrow$ original parity-check matrix of the code\\
$\mathrm{\textbf{H}^e} \Leftarrow$ a set of parity-check equations not in $\mathrm{\textbf{H}}$\\
$\mathcal{T} \Leftarrow$ number of trials\\
$\mathcal{\psi} \Leftarrow$ number of selections
\ENSURE a codeword with the minimum euclidean distance
\STATE Perform BP decoding and $\mathbf{d}_0 \Leftarrow$ decoded output.
\STATE $\mathrm{d_E(\mathbf{d}_0, \mathbf{r})} \Leftarrow$ euclidean distance between $\mathbf{d}_0$ and $\mathbf{r}$.
\STATE $\mathbf{d^\prime} \Leftarrow \mathbf{d}_0$ and $\mathrm{d_E^{min}} \Leftarrow \mathrm{d_E(\mathbf{d}_0, \mathbf{r})}$
\FOR {$\mathcal{\tau} = 1$ to $\mathcal{T}$,}
\FOR {$i = 1$ to maximum number of iterations,}
\STATE Pick $\mathcal{\psi}$ parity-check equations from $\mathrm{\textbf{H}^e}$.
\STATE Substitute them into $\mathrm{\textbf{H}}$ to generate $\hat{\mathrm{\textbf{H}}}$.
\STATE Based on $\hat{\mathrm{\textbf{H}}}$, perform the check nodes (horizontal) and bit nodes (vertical) processing as in standard BP algorithm,
\STATE $\mathbf{d}_{\mathcal{\tau}} \Leftarrow$ denote the decoded output,
\STATE $\mathrm{d_E(\mathbf{d}_\mathcal{\tau}, \mathbf{r})} \Leftarrow$ euclidean distance between $\mathbf{d}_{\mathcal{\tau}}$ and $\mathbf{r}$.
\IF {$\mathbf{d}_{\mathcal{\tau}}\mathrm{\textbf{H}}^T = \mathbf{0}$}
\STATE Stop the algorithm
\ENDIF 
\ENDFOR
\IF {$\left(\mathrm{d_E(\mathbf{d}_\mathcal{\tau}, \mathbf{r})} < \mathrm{d_E^{min}}\right)$ and
$\left(\mathbf{d}_{\mathcal{\tau}}\mathrm{\textbf{H}}^T = \mathbf{0}\right)$}
\STATE $\mathbf{d^\prime} \Leftarrow \mathbf{d}_{\mathcal{\tau}}$ and $\mathrm{d_E^{min}} \Leftarrow \mathrm{d_E(\mathbf{d}_\mathcal{\tau}, \mathbf{r})}$
\ENDIF
\ENDFOR
\STATE Output $\mathbf{d^\prime}$.
\end{algorithmic}
\end{algorithm}

\section{Simulation Results}\label{sec:results}
In this section, we present simulation results of the modified BP decoder for
some cyclic codes designed using the approach discussed in section~\ref{sec:fft-cyclic}.
The selection of the parity-check equations is made on a random basis.
It is assumed that the simulation system employs BPSK modulation mapping
the symbols 0 and 1 to $-1$ and $+1$ respectively. 

\begin{figure}[t]
\centering
\includegraphics[width=2.4in,angle=270]{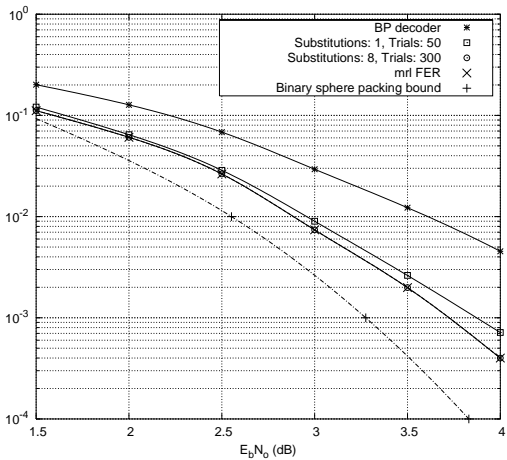}
\caption{\label{fig:fer_63_37}FER performance of the $(63,37)$ cyclic code.}
\end{figure}

Fig.~\ref{fig:fer_63_37} shows the FER performance of the $(63,37)$ cyclic code.
It is shown that the modified
BP decoder, provided enough substitutions and trials are used, can achieve the ML performance
as indicated by the mrl-FER and the FER of the modified BP decoder that produce the same curve.
Compared to the standard
BP decoder, at a FER of $10^{-3}$ a gain of approximately $0.9$ dB is obtained by using
the modified BP decoder. In addition, it can be seen that, at a FER of $10^{-3}$, the performance
of the code is within $0.4$ dB of the sphere-packing-bound~\cite{Shannon.1959},\cite{Dolinar_et_al.1998}
for a code of length $63$ and code-rate of $0.587$ after allowing for the coding loss attributable to binary
transmission. Table~\ref{tbl:mrlpercentage_63_37} shows how close the performance of the
modified BP decoder is to the ML decoder. With the standard BP decoder more than
$50\%$ mrl codewords are found in the low signal-to-noise ratio (SNR) region, but in the moderate SNR
region we can only find a few mrl codewords. With just single substitution and 50 trials,
the modified BP decoder is able to increase the percentage of mrl codewords
found to be higher than $50\%$. ML performance is achieved with 8 substitutions and 300 trials. 

\begin{figure}[t]
\centering
\includegraphics[width=2.4in,angle=270]{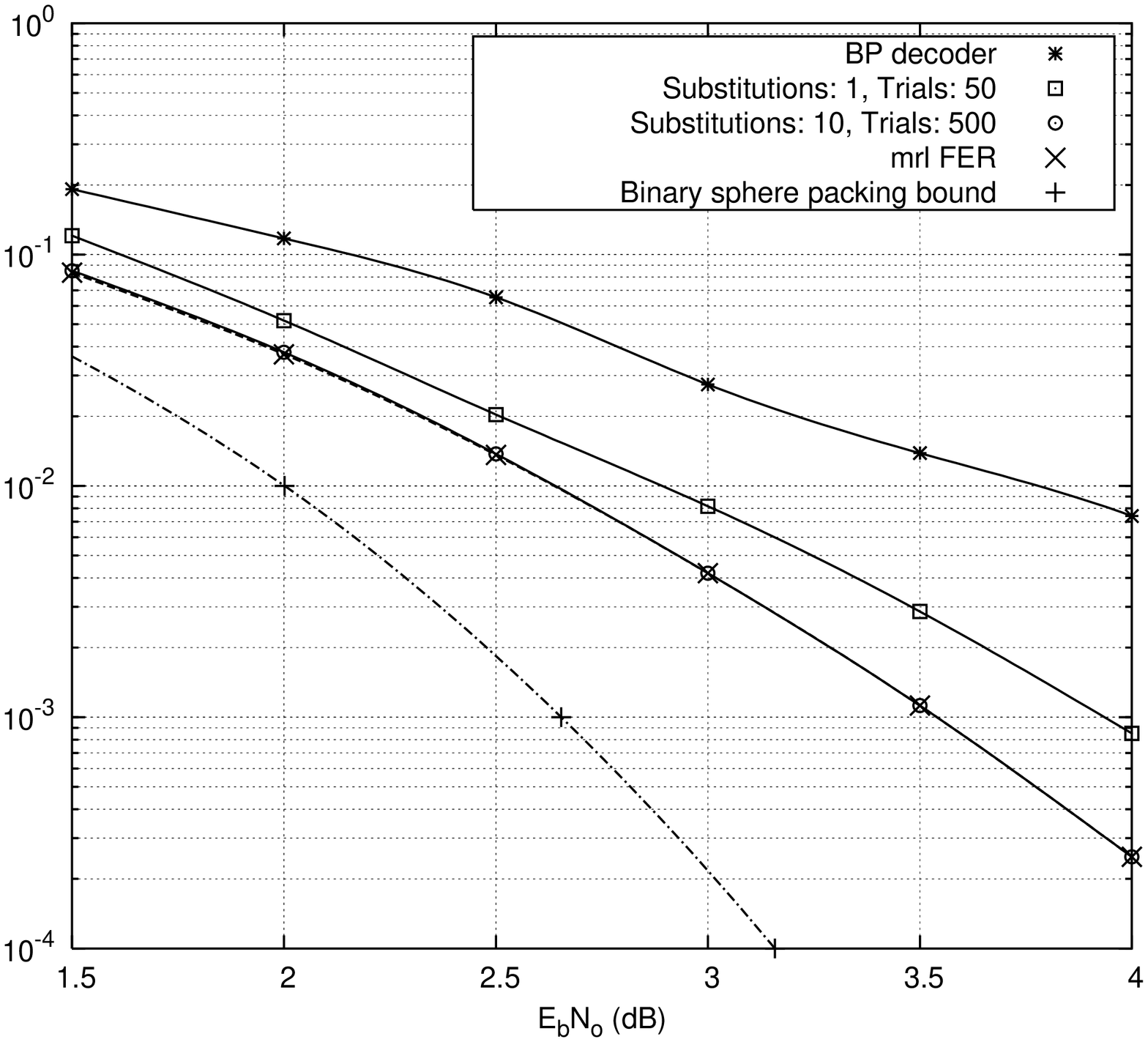}
\caption{\label{fig:fer_93_47}FER performance of the $(93,47)$ cyclic code.}
\end{figure}
\begin{figure}[t]
\centering
\includegraphics[width=2.4in,angle=270]{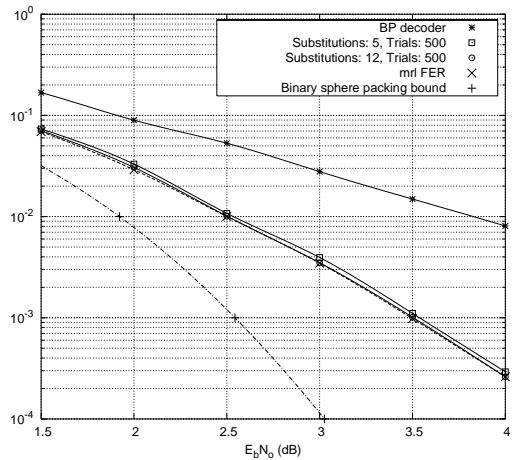}
\caption{\label{fig:fer_105_53}FER performance of the $(105,53)$ cyclic code.}
\end{figure}
\begin{figure}[t]
\centering
\includegraphics[width=2.4in,angle=270]{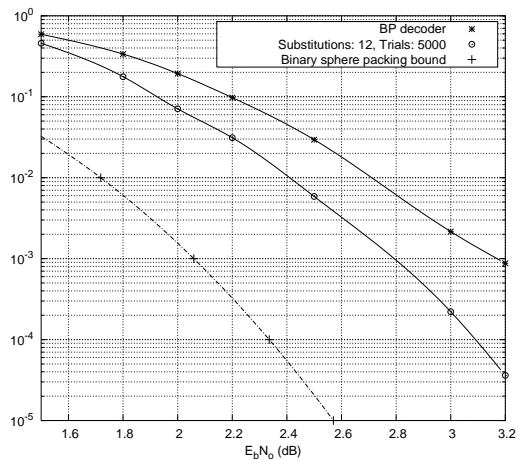}
\caption{\label{fig:fer_341_205}FER performance of the $(341,205)$ cyclic code.}
\end{figure}

\begin{table}[hbt]
\caption{\label{tbl:mrlpercentage_63_37}Percentage of mrl codewords against $E_b/N_o$($\mathrm{dB}$) of the $(63,37)$ cyclic code}
\centering
\begin{tabular}{|c|c|c|c|c|c|c|}\hline\hline
\multicolumn{7}{|c|}{Standard BP decoder}\\\hline
$E_b/N_o$ & $1.5$ & $2.0$ & $2.5$ & $3.0$ & $3.5$ & $4.0$\\\hline
$\%$ & $73$ & $41$ & $27$ & $23$ & $16$ & $9$\\\hline\hline
\multicolumn{7}{|c|}{Substitutions: $1$, Trials: $50$}\\\hline
$E_b/N_o$ & $1.5$ & $2.0$ & $2.5$ & $3.0$ & $3.5$ & $4.0$\\\hline
$\%$ & $90$ & $94$ & $90$ & $82$ & $74$ & $61$\\\hline\hline
\multicolumn{7}{|c|}{Substitutions: $8$, Trials: $300$}\\\hline
$E_b/N_o$ & $1.5$ & $2.0$ & $2.5$ & $3.0$ & $3.5$ & $4.0$\\\hline
$\%$ & $100$ & $100$ & $100$ & $100$ & $100$ & $100$\\\hline\hline
\end{tabular}
\end{table}
\begin{table}[hbt]
\caption{\label{tbl:improvement}Performance gain with the respect to BP decoder and
distance from sphere-packing-bound$^\dagger$ at the FER of $10^{-3}$}
\centering
\begin{tabular}{|c|p{1.1in}|p{1.3in}|}\hline\hline
Codes & Gain with the respect to BP decoder & Distance from sphere-packing-bound$^\dagger$\\\hline\hline 
$(63,37)$  & \multicolumn{1}{|c|}{$0.9$ dB} & \multicolumn{1}{c|}{$0.4$ dB}\\
$(93,47)$  & \multicolumn{1}{|c|}{$1.1$ dB} & \multicolumn{1}{c|}{$0.8$ dB}\\
$(105,53)$ & \multicolumn{1}{|c|}{$2.0$ dB} & \multicolumn{1}{c|}{$0.9$ dB}\\
$(341,205)$& \multicolumn{1}{|c|}{$0.4$ dB} & \multicolumn{1}{c|}{$0.7$ dB}\\\hline\hline
\end{tabular}
\begin{flushleft}
{\scriptsize $^\dagger$ Sphere-packing-bound offset by binary transmission loss.}
\end{flushleft}
\end{table}

Fig.~\ref{fig:fer_93_47}, \ref{fig:fer_105_53} and \ref{fig:fer_341_205} show the FER
performance of the $(93,47)$, $(105,53)$ and $(341,205)$ cyclic codes respectively. Both of the
$(93,47)$ and $(105,53)$ codes achieve ML performance with the modified BP decoder. For
the $(341,205)$ code, the modified BP decoder produces a gain of approximately $0.4$dB
with the respect to the standard BP decoder. Due to the code length, there are very few
mrl codewords observed indicating that a better decoder is required.
Table~\ref{tbl:improvement} summarises, at the FER of $10^{-3}$,
the amount of gain obtained with the modified decoder with the respect to the standard BP
decoder and the distance from the sphere packing bound after allowing binary transmission
loss. 

\section{Conclusions}\label{sec:conclusion}
Construction of cyclic LDPC codes using idempotents and MS polynomials  can produce a
large number of cyclic codes that are free from cycle of length 4. Some of the codes are
already known such as the DSC codes, but others are new. An important feature of this
approach is the ability to increase the $d_{min}$ of the codes by taking into account
additional irreducible factors of $1+z^n$ and so steadily decrease the sparseness of the
parity-check matrix. As an example, consider that we want to design a cyclic code of length
$63$. If we let $f(z) = 1+z+z^6$, we obtain a $(63,31)$ cyclic code with $d_{min}$ of 7.
Now, if the irreducible polynomial $1+z+z^2$ is also taken into account, the resulting cyclic
code is the $(63,37)$ code which has $d_{min}$ of 9. The row or column weight of the
parity-check matrices for the former and latter codes are $6$ and $8$ respectively.

By substituting the parity-check equations in the parity-check matrix with other codewords
of the dual code derived from their linear combinations, the performance of the BP decoder
can be improved. For the $(63,37)$, $(93,47)$ and $(105,53)$ cyclic codes, the modified BP
decoder has been shown to achieve ML performance.

Although the substitution method introduces cycles of length 4, these cycles do not pose a
lasting negative effect on the iterative decoder. By substituting at every iteration, the effect of these
short cycles is broken and simulation results have shown that this can improve the decoding
performance.


\end{document}